# مدلسازی اپیدمی و هموارسازی منحنی همه گیری در شبکه های انسانی-اجتماعی

محمدرضا دوست محمدیان[1]،*، ثریا دوست محمدیان[2]، نجمه دوست محمدیان[3]، اعظم دوست محمدیان[4]، هومن ضرابی[5] و حمیدرضا ربیعی[6]

**چکیده**

در این مقاله، مساله مدلسازی اپیدمی و هموار کردن منحنی همه گیری بیماریها در شبکه های انسانی-اجتماعی مورد بررسی قرار میگیرد. هموارتر کردن نمودار همه گیری به معنی کند کردن گسترش بیماری و کاهش نرخ انتقال است که با استفاده از فاصله گذاری اجتماعی، ایزوله کردن افراد و البته واکسیناسیون انجام می شود. روشهای غیردرمانی البته راههای ساده تر و سریعتری برای کنترل نرخ گسترش و اپیدمی بیماری هستند. با هدفمندتر کردن این روشهای غیر درمانی برای گروههایی مشخص با مرکزیت بالاتر در ساختار جامعه میتوان به نسبت نمودار هموارتری برای همه گیری بیماری مثل کرونا داشت بدون اینکه هزینه های درمانی خاصی تحمیل گردد. هدف در این پژوهش ابتدا مدلسازی مساله اپیدمی و سپس ارایه راهکارها و الگوریتمهای ساختاری بر مبنای ساختار شبکه انسانی اجتماعی به منظور واکسیناسیون هدفمندتر یا روشهای غیردرمانی هدفمندتر برای کاهش پیک بیماری واگیر و هموارکردن منحنی همه گیری می باشد. این راهکارها براساس ساختار گراف شبکه انسانی-اجتماعی بوده و میتوانند تا حد محسوسی در کاهش نرخ انتقال موثر باشند. بدین منظور تعداد خاصی از نودهای شبکه با مرکزیت بالا ایزوله شده و سپس نمودار همه گیری شبکه بررسی می شود. این تحقیق نتایج معناداری برای هموارکردن نمودار همه گیری شبکه تنها با ایزوله کردن درصد کمی از نودهای خاص را نشان می دهد. روشهای ارایه شده در این تحقیق مستقل از نوع بیماری بوده و برای انواع بیماریهای واگیردار از جمله کووید-۱۹ موثر است.

## ۱-مقدمه

همه گیری و بیماریهای واگیردار هزینه های زیادی به کشور تحمیل میکنند. این در حالی است که با روشهای پیشگیرانه هدفمند تا حد زیادی میتوان از این هزینه ها بر کادر درمان و پزشکی کاست. به طوری که به نقل از وزارت بهداشت همه گیری کرونا تاکنون ۴۲ هزار میلیارد تومان هزینه تنها به نظام سلامت کشور تحمیل کرده است. بدیهی است با روشهای پیشگیرانه و واکسیناسیون هدفمند میتوان درصد بالایی از این هزینه ها را که ممکن است در مواجهه پیکها و سویه های جدید کرونا تحمیل شوند از دوش دولت برداشت. در این بین روشهای غیر دارویی (non-pharmaceutical) هزینه بسیار کمتری از روشهای دارویی (pharmaceutical) به نظام درمان کشور تحمیل میکنند. این روشها شامل پوشیدن ماسک، استفاده از الکل و شستن دستها، و البته ایزوله کردن افراد و فاصله گذاری اجتماعی به عنوان مهمترین روش غیردارویی میشوند. این روشهای غیردارویی ساده نه تنها هزینه چندانی به نظام درمان تحمیل نمیکنند بلکه با کاهش پیک جمعیتی افراد بیمار بستری و همچنین به

* doost@semnan.ac.ir
۱. استادیار، دانشکده مهندسی مکانیک، دانشگاه سمنان
۲. استادیار، دانشگاه علوم پزشکی سمنان
۳. استادیار، دانشگاه علوم پزشکی سمنان
۴. استادیار، مرکز تحقیقات بیماریهای گوارش و کبد، دانشگاه علوم پزشکی ایران
۵. استادیار، پژوهشگاه مخابرات ایران (ITRC)
۶. استاد، دانشکده مهندسی کامپیوتر، دانشگاه صنعتی شریف

اصطلاح پزشکی تخت کردن منحنی همه گیری (flattening the infection curve) هزینه های وارد بر کادر درمان، بیمارستانها و نظام سلامت را تا حد زیادی کاهش میدهد.

فاصله گذاری اجتماعی و قرنطینه های فردی در کنار رعایت دیگر پروتوکل های بهداشتی از راههای اصلی جلوگیری و کاهش گسترش ویروس کووید-۱۹ و بیماری کرونا می باشد. از طرف دیگر واکسیناسیون عمومی از مهمترین راهکارهای دارویی و پزشکی برای جلوگیری از این بیماری می باشد. در سالهای اخیر این روشها دو راهکار اصلی فردی در کنار قرنطینه های عمومی برای مبارزه با گسترش این ویروس در کشورهای مختلف بوده است. تحقیقات نشان میدهد در این بین قرنطینه و واکسیناسیون "هدفمند" تا حد محسوسی در کنترل این بیماری و کاهش موارد بستری در بیمارستانها موثر بوده است. یک دلیل این استراتژی های هدفمند کاهش تعداد موارد مثبت در پیک بیماری و کم کردن بار تحمیلی بر سیستم درمان میباشد. به این شکل سیستم و کادر درمانی و بیمارستانها به شکل بهتری قادر به ادامه کار و خدمت رسانی به بیماران خواهند بود .

در این پژوهش هدف ارایه مدل کردن مساله اپیدمی و سپس ارایه استراتژی های موثرتر قرنطینه و واکسیناسیون "هدفمند" برای کاهش پیک جمعیتی افراد بیمار بستری و به اصطلاح کشیده تر کردن یا هموارکردن منحنی همه گیری (flattening the infection curve) [1] با استفاده از مطالعه ساختاری اپیدمی و همه گیری در شبکه های انسانی می باشد. از آنجا که در بسیاری از موارد قرنطینه عمومی هزینه های بالای اقتصادی به برخی کسب و کارها وارد میکند قرنطینه های فردی با توجه به موقعیت شغلی و اجتماعی افراد در ساختار جامعه و به اصطلاح مرکزیت اجتماعی (social centrality) میتواند اقدام جایگزین موثری باشد. نکته مهم تعیین مرکزیت افراد به صورت شخصی و محلی است بطوری که هر شخص براساس مناسبات اجتماعی و تماسهای خود با دیگران قادر به تشخیص مرکزیت اجتماعی نسبی خود بوده و بتواند با قرنطینه فردی و شخصی از افزایش نرخ انتقال بیماری جلوگیری کند. به دلیل مشابه واکسیناسیون هدفمند براساس موقعیت اجتماعی افراد در جامعه میتواند به طریق موثرتری نرخ انتقال ویروس را کند کرده و پیکهای بیماری را کاهش دهد. استخراج چنین موقعیتهای اجتماعی براساس ساختار شبکه انسانی براساس تحلیل داده های اپیدمی موجود میتواند تا حد مناسبی به این مساله کمک کنند و ازنظر اقتصادی تاثیرات منفی کمتری بر کسب و کارها، خدمات عمومی و خدمات بهداشتی درمانی دارد. این استراتژی ها  نه تنها برای همه گیری بیماری کرونا بلکه همه گیری بیماریهای ویروسی مشابه را شامل می شود.

**پیشینه تحقیق:** بررسی و مطالعه داده های پزشکی و همه گیری به همراه ارایه روشهای پیشگیرانه از قبیل ایزوله کردن، فاصله گذاری اجتماعی، واکسیناسیون برای پیشگیری از پیکهای سویه های جدید کرونا را بیش از پیش مورد توجه قرار داده است. اکثر این روشهای پیشگیرانه به منظور کم کردن نرخ انتقال بیماری، کاهش پیک بیماری (بطور مثال در همه گیری بیماری کرونا) و به اصطلاح هموارکردن منحنی همه گیری بیماری (flattening the infection curve) صورت میگیرد [1,2]. به این ترتیب که با شبیه سازی شبکه های انسانی-اجتماعی به صورت گرافی شامل نودهایی که نمایانگر افراد جامعه هستند و تماس افراد و راههای انتقال بیماری به صورت یالهای گراف به مطالعه گسترش ویروس کرونا از یک یا چند نود آلوده به سایر نودها پرداخته می شود. هر چه توزیع نودهای آلوده در طی زمان هموارتر و کشیده تر باشد نشان دهنده کم بودن بار پیک بیماری در طی زمان می باشد [2,3,4,5,6,7].

ترکیبی از مدلهای آماری و احتمالاتی با نام مدلهای بخش بندی (compartmental model) [8,9,10] به همراه ساختار شبکه انسانی-اجتماعی (social network) برای تحلیل گسترش اپیدمی و نحوه قرنطینه و بازگشایی قرنطینه ها در ادبیات مساله مورد بررسی قرار گرفته است. انواع مدلهای احتمالاتی موجود با درنظر گرفتن مواردی همچون ایمن شدن نسبت به ویروس پس از بیماری، مدت زمان ایمنی پس از واکسیناسیون، میزان واگیردار بودن بیماری در صورت تماس و نوع تماس طبقه بندی می شوند [3,4]. این مدلها محدود به بیماری کرونا نمی شوند بلکه سایر بیماریها مثل آنفلوانزا را هم شامل می شوند [11,12,13,14]. توجه شود که مدل کردن تعاملات انسانی براساس گرافهای واقعی [15,16] و رندوم [17,18,19] و سپس آنالیز گسترش اپیدمی یا سایر انواع

تغییرات ساختاری بسیار معمول و رایج است. بطور مثال در مرجع [15] با استفاده از موقعیت افراد و تماسهایی که با هم داشته اند یک مدل واقعی برای شبکه های اجتماعی ارایه شده است که مشابهتهایی را با مدلهای رندوم [17,18,19] موجود نشان می دهد. همچنین مدلهای احتمالاتی بخش بندی بسیاری برای مدل سازی نرخ انتقال بیماری بین افراد آلوده و غیر آلوده ارایه شده است [1,8,9,10]. یادآوری می شود که مدل‌های بخش‌بندی مدل‌سازی ریاضی بیماری‌های واگیردار را ساده می‌کنند. در این مدل‌ها، جمعیت مورد مطالعه به بخش‌های محدودی با برچسب‌هایی مثل S مستعد بیماری، I آلوده یا بیمار تقسیم‌بندی می‌شوند. اعضای جامعه می‌توانند از بخشی به بخش دیگر بروند وضعیتشان تغییر کند. ترتیب این برچسب‌ها در نام‌گذاری این مدل‌ها نشان‌دهنده جریان تغییر وضعیت از بخشی به بخش دیگر است [1].

مطالعات اخیر روابط معناداری بین گسترش همه گیری (بطور مثال اپیدمی کرونا) و ساختار و خصوصیات گراف شبکه انسانی و اجتماعی که ویروس در آنها گسترش می یابد را نشان میدهد. مطالعات و تحقیقات اکثرا از دید علم شبکه های پیچیده (complex networks) و تئوری گرافها (graph theory) [17,18,19] همچنین روابط و مشابهتهای معناداری بین گسترش بیماری ها در شبکه های انسانی و گسترش ویروسها (وهمچنین بدافزارها) در شبکه های اجتماعی سایبری را نشان میدهند. این مشابهتها مربوط به ساختار شبکه ها بوده و در عمل تحلیلهای آماری و معادلات ریاضی مشابهی بر آنها حاکم است.

در مقایسه با کارهای قبلی [1,2,12] که از روش ایزوله کردن (حذف کردن) لینکها برای هموار کردن منحنی اپیدمی استفاده می کنند در این مقاله از روش ایزوله کردن نودها استفاده شده است. به عبارت دیگر، این مقاله واکسیناسیون یا قرنطینه افراد را شبیه سازی میکند ولی مقاله [1,2,12] فاصله گذاری اجتماعی را شبیه سازی می کنند. در مقایسه با مدلهای احتمالاتی استفاده شده در [4,8,9] از مدل مستعد-آلوده (susceptible-infected) در این مقاله استفاده شده است به این معنی که افراد جامعه یا در حالت مستعد بیماری هستند یا آلوده به بیماری هستند. در مقایسه با مدل شبکه بی مقیاس در

[18] که قابلیت تغییر خوشه بندی شبکه را ندارد از مدل شبکه با خوشه بندی متغیر در این مقاله استفاده شده است تا اثر خوشه بندی بر هموارسازی منحنی اپیدمی بهتر مشخص شود.

## 2- مدل کردن اپیدمی در شبکه ها

### 2-1- مدل کردن شبکه های انسانی-اجتماعی

شبکه به شکل یک گراف با تعداد مشخصی نود (یا گره) و تعدادی لینک (یا یال) که ارتباط دهنده نودها هستند شبیه سازی میشوند. در گراف شبکه های انسانی-اجتماعی هر نود نماینده یک شخص و هر لینک بین دو نود نماینده ارتباط یا اثرگذاری بین دو شخص می باشد. اثرگذاری دو نود میتواند جهتدار یا بدون جهت (دوطرفه) باشد. در لینکهای جهتدار اثرگذاری از سمت یک شخص روی شخص دیگر است (به این معنی که تنها یک شخص روی شخص دیگر اثر میگذارد) و در لینکهای دوطرفه اثرگذاری و ارتباط دو شخص دوطرفه میباشد و از سمت هردوشخص می باشد.

گرافهای رندوم (یا تصادفی) مدل های ریاضی هستند که لینک بین نودها را به شکل رندوم مدل میکنند و معروفترین این گرافها برای مدل کردن ارتباطات بین افراد جامعه در شبکه های انسانی-اجتماعی با نام شبکه های بی مقیاس (scale-free) شناخته می شوند. در ساختار این نوع گرافها، لینکها با روشهای احتمالاتی بین نودها برقرار می شوند بگونه ای که توزیع درجه نودها (یا تعداد لینکهای متصل به هرنود) تحت تبدیل مقیاس، بدون تغییر باقی میماند. بی‌مقیاس بودن در این گونه شبکه‌ها به این معناست که با چند برابر کردن متغیر توزیع درجه نودها (مرتبط با احتمال برقراری لینکها)، شکل توزیع تغییری نمی کند. نحوه اتصال نودها در یک ساختار بی‌مقیاس براساس اتصال ترجیحی (preferential attachment) می باشد. معروفترین مدل برای ساخت تصادفی چنین گرافهایی مدل باراباشی-آلبرت می باشد [18]. در این مدل احتمال ارتباط یک نود با نودهای دیگر براساس درجه نود می باشد. توجه شود که درجه نود براساس تعداد لینکهای متصل به نود تعیین می شود و دو نود که با هم مرتبط هستند همسایه نامیده می شوند. به این ترتیب نودهای با درجه بیشتر (یا تعداد همسایه های

بیشتر) با احتمال قویتری به نود جدید وصل می شوند. به عبارت دیگر نودهای با درجه بیشتر دارای مرکزیت بالاتری در گراف هستند. در شبکه های انسانی این اتصال ترجیحی به این معنی است که افراد جدید ترجیح می دهند با افراد قدیمی که تعداد دوستان بیشتری دارند دوست شوند. در مدل هولم-کیم نمونه‌ای دیگر از ساز و کارهای پیشنهادی برای ایجاد شبکه‌های بی‌مقیاس است [19]. در این مدل علاوه بر اتصال ترجیحی افزایش خوشه بندی (clustering) شبکه نیز مدنظر است. به این منظور علاوه بر اتصال ترجیحی نودها تمایل به اتصال سه تایی یا مثلث (triad) دارند. به عبارت دیگر هر نود علاوه بر اتصال به نودهای دارای مرکزیت و درجه بالاتر به یک یا چندتا از همسایه های این نودها نیز متصل می شود. این اطلاعات در شکل (1) به اختصار توضیح داده شده است. در شبکه انسانی این روش به این معنی است که نودها پس از ارتباط با افراد با تعداد دوستان بالا ترجیحا با یک یا چند دوست فرد مذکور نیز دوست می شوند.

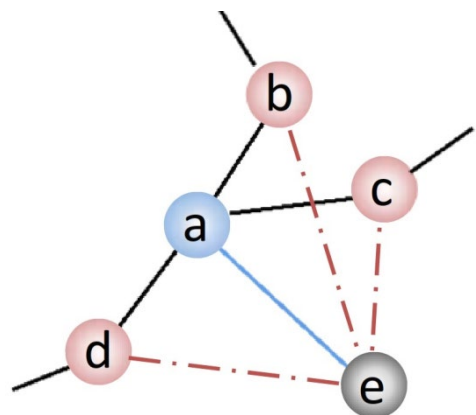

شکل 1- احتمال اتصال نود جدیدمشکی e به نودهای موجود متناسب با درجه نودهاست. با توجه به اینکه درجه نودهای قرمز b,c,d برابر 2 و درجه نود آبی a برابر 3 است ترجیح (و احتمال) اتصال نود مشکی به نود آبی 1.5 برابر نودهای قرمز است. این اصل با عنوان اتصال ترجیحی شناخته می شود. برای افزایش خوشه بندی پس از اتصال نود مشکی به نود آبی با درجه بالا، نود جدید مشکی به یکی از همسایه های قرمز این نود نیز متصل می شود تا یک مثلث یا سه تایی تشکیل شود. به این اتصال تشکیل سه گانه (triad) گفته می شود [20].

مدل هولم-کیم بیشتر به شبکه‌های انسانی-اجتماعی واقعی نزدیک است چرا که بر خلاف مدل باراباشی-آلبرت،

مدل هولم-کیم شبکه‌هایی با ضریب خوشگی (clustering coefficient) بالاتری تولید می‌کند. ضریب خوشگی در یک شبکه معیاری است که درجه گره‌ها در یک گراف که تمایل به ایجاد یک خوشه با هم دارند را اندازه می‌گیرد. ضریب خوشگی سراسری بر پایه یک سه تایی از گره‌ها تعریف می‌شود. یک سه تایی متشکل از سه گره متصل به هم است و بنابراین یک مثلث شامل سه سه‌تایی است که هر‌یک به مرکزیت یکی از گره ها است. ضریب خوشگی نسبت تعداد کل سه‌تایی‌های بسته (یا سه برابر تعداد کل مثلث ها) به تعداد کل سه‌تایی هاست (سه‌تایی‌های باز و بسته). در یک شبکه انسانی دارای مفهوم مشخصی است. به این معنی که در شبکه با ضریب خوشگی بالا اگر دو شخص دارای یک همسایه (یا دوست) مشترک باشند به احتمال بیشتری با یکدیگر همسایه (دوست) هستند و بالعکس در شبکه انسانی با ضریب خوشگی پایین دو شخص با همسایه (دوست) مشترک با احتمال کمتری در ارتباط هستند. در مورد ضریب خوشگی مراجع [21] و [22] جزییات بیشتر و اطلاعات مناسبتری می دهند.

## 2-2- مدل کردن اپیدمی

برای مدل کردن اپیدمی از مدل شبکه بی مقیاس بدون جهت (با ارتباطات دوطرفه) برای شبکه انسانی-اجتماعی استفاده شده است. هدف مدل کردن نحوه انتقال آلودگی نودها (بیماری یا ویروس) از یک نود آلوده مرجع است. در هر پله زمانی نود آلوده تمام نودهای همسایه خود را آلوده میکند و سپس نودهای همسایه آلده شده همسایه های خود را آلوده میکنند و به این ترتیب بیماری پله به پله از یک نود آلوده در طی زمان به نودهای دیگر منتقل می شود. رابطه بین زمان آلوده شدن نود های جدید نسبت به نود مرجع با مفهوم کوتاه ترین مسیر در گراف (shortest path) ارتباط دارد. مسیر بین دو نود در یک گراف به معنی یال‌های متصل کننده دو نود در گراف هستند و طول این مسیر تعداد یال‌هایی است که باید از آن عبور کرد تا از یک نود به نود دیگر رسید. به عبارت دیگر دنباله ای از نودهای دوبه دو متمایز که از نود u شروع و به نود v ختم می‌شود به طوری که هر دو متوالی این دنباله در گراف مجاور یا همسایه هم باشند. طول یک مسیر برابر است با تعداد یال‌های موجود در آن مسیر (یکی کمتر از

تعداد رئوس موجود در آن مسیر. مسیر بین دو نود به طور کلی یکتا نیست و کوتاه ترین این مجموعه یالها (یا مجموعه نودها) به عنوان کوتاه ترین مسیر شناخته میشود.

در ادامه دو مثال ساده برای روشن شدن بهتر مساله و راهکار پیشنهادی آمده است. در شکل (2) با داشتن گراف یک شبکه انسانی-اجتماعی منحنی همه گیری شبکه براساس کوتاه ترین مسیر (فاصله) نودها از نود آلوده مشخص شده است. نودهای همرنگ دارای فاصله یکسان (فاصله از لحاظ تعداد یالها) از نود آلوده به رنگ مشکی هستند. این نودهای همرنگ در فاصله زمانی یکسانی نسبت به نود آلوده قرار دارند و با آلوده شدن این نودها نمودار همه گیری براساس تعداد نودهای آلوده شده در فاصله یکسان از نود آلوده مرجع قابل رسم است. در واقع نودهای همرنگ در کوتاه ترین مسیر یکسانی از نود مرجع یا ریشه قرار دارند. در نمودار میله ای رنگ هر میله براساس رنگ نودها در شبکه مشخص شده است. با فرض اینکه شبکه نودها شبکه بیمقیاس scale-free باشد توزیع نمودار میله ای کوتاه ترین مسیر نودها به شکل توزیع گاما تابع گاما خواهد بود. با پوشاندن (fit) توزیع گاما به نمودار میله ای کوتاه ترین مسیر نودها می توان نمودار همه گیری (infection curve) شبکه را یافت. در صورت بالاتر بودن این منحنی از خط فرضی ظرفیت مراکز خدماتی بهداشتی درمانی (خط نقطه چین در شکل) فشار زیادی به این مراکز برای پوشش و درمان بیماران وارد می شود.

مدل موجود در این مقاله بر اساس تیوری گرافها و مدل بخش‌بندی (compartmental) [1] برای اپیدمی استفاده شده. مدل آسیب‌پذیر- آلوده یا مستعد-بیمار (susceptible-infected) که در این مقاله استفاده شده یکی از مدلهای معروف در اپیدمیولوژی است که البته مشابهتهایی هم با مدلهای گسترش ویروس در فضای مجازی دارد. در مدل آسیب‌پذیر- آلوده استفاده شده نرخ انتقال بسیار بالا در نظر گرفته شده به این معنی که تمام همسایه های نود آلوده در مرحله بعدی با نرخ احتمال 1 آلوده می شوند و به این ترتیب نرخ انتقال در مدل آسیب‌پذیر- آلوده نزدیک 1 درنظر گرفته شده است.

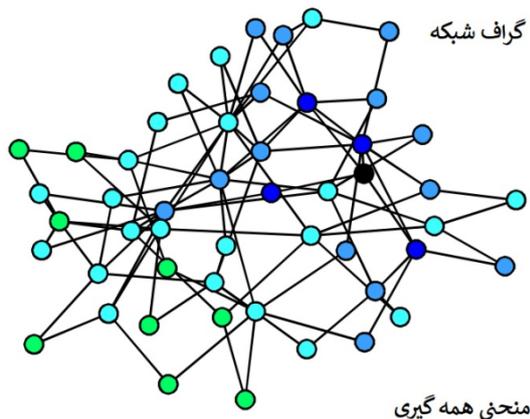

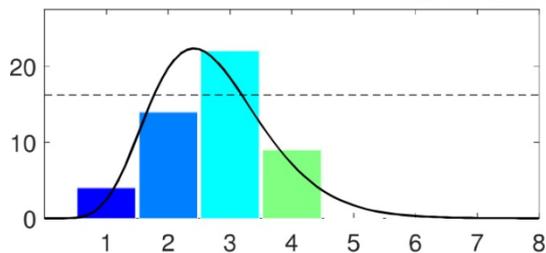

شکل 2- انتقال بیماری از نود آلوده (به رنگ مشکی) به نودهای همجوار. نودهای همرنگ دارای فاصله ساختاری مشابهی (از دید گراف شبکه) از نود آلوده هستند.

یک مثال ساده برای هموار کردن این منحنی همه گیری ایزوله کردن تعدادی از نودهاست. بطور مثال با قرنطینه هدفمند تنها 3 نود قرمز (6 درصد افراد جامعه هدف) میتوان ارتباط این نودها با سایر نودهای شبکه را قطع کرد. این موضوع با حذف یالهای نودهای قرمز در شکل (3) نشان داده شده است که باعث افزایش طول کوتاه ترین مسیر بین اکثر نودهای شبکه با نود آلوده میشود. به این ترتیب نودهای بیشتری در فاصله دورتری (کوتاه ترین مسیر بزرگتر) از نود آلوده قرار می گیرند و منحنی انتشار ویروس در شکل (3) تا حد محسوسی کشیده تر و هموارتر شده و مسیر انتقال ویروس از نود (شخص) آلوده به سایر نودها (اشخاص جامعه) تا حد محسوسی طولانی تر شده است. به این ترتیب پیک منحنی انتقال ویروس یا همان منحنی همه گیری (infection curve) کاهش یافته و در زیر خط فرضی نشانگر ظرفیت مراکز خدماتی بهداشتی درمانی قرار گرفته است.

(نودهای مجاور) یک نود. برای نود i درجه نود با d(i) مشخص می‌شود.

**نزدیکی (closeness):** برابر است با عکس فاصله یا کوتاه‌ترین مسیر یک نود تا نودهای دیگر. نودی که بیشترین مقدار نزدیکی را دارد سرعت دسترسی بیشتری به نودهای دیگر دارد و میتواند در مدت زمان کمی به همه نودها اطلاعات بدهد یا (در این مقاله) بیماری را منتقل کند.

**بینابینی (betweenness):**
برابر است با نسبت تعداد دفعاتی که یک نود روی کوتاه‌ترین مسیر نودهای دیگر قرار می‌گیرد. در واقع این معیار اهمیت نسبی یک نود را با استفاده از ترافیکی که آن نود بین بقیه نودها ایجاد کرده بیان می‌کند که این مقدار برابر است با تقسیم تعداد مسیرها بین تمام جفت نودها بر تمام مسیرها بین جفت نودهایی که نود مورد نظر را دربردارند. در واقع این معیار مشخص میکند که چه تعداد از نودهای شبکه برای ارتباط سریعتر باهم (با واسطه کمتر) به این نود نیاز دارند.

**بوناچیچ (Bonacich):** این معیار E(i) با عنوان بردار ویژه (eigen centrality) هم شناخته می‌شود و مقدار آن با استفاده از ماتریس مجاورت (adjacency matrix) شبکه بدست می‌آید [۲۴]. در واقع E بردار ویژه چپ مربوط به بزرگترین مقدار ویژه ماتریس مجاورت است. اگر نودی به نودهای با مرکزیت بالا متصل باشد تحت تاثیر آن اهمیت یا مرکزیت نود مذکور نیز بالا می‌رود. این معیار در واقع با روشی تکراری مرکزیت نودهای مجاور یا همسایه را نیز درنظر می‌گیرد.

**کاتز (Katz):** طبق این معیار یک نود ارزش بالایی دارد اگر هم خودش درجه بالایی داشته باشد و به نودهای با درجه بالا مرتبط باشد و آن نودهای مجاور خود به نودهای با درجه بالا متصل باشند و به همین ترتیب ارتباط با نودهای دورتر اثرگذار است. در این معیار نودهای در فاصله دورتر با یک ضریب مثلا a شمرده می‌شوند. به عبارت دیگر از جمع تمام مسیرهای ژئودسی (geodesic)

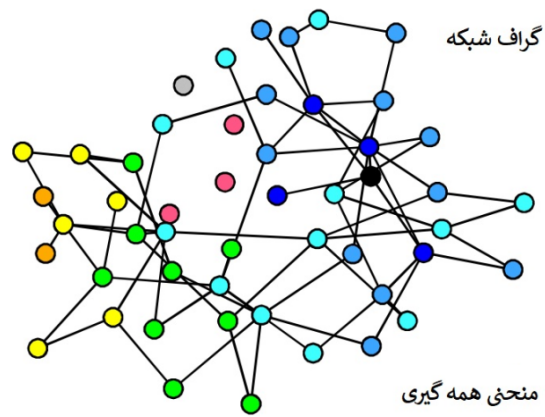

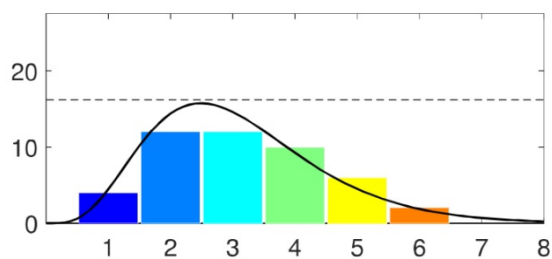

شکل ۳- انتقال بیماری از نود آلوده (به رنگ مشکی) به نودهای همجوار. نودهای همرنگ دارای فاصله ساختاری مشابهی (از دید گراف شبکه) از نود آلوده هستند. با ایزوله کردن تنها سه نود قرمز (با مرکزیت بالا در شبکه) منحنی همه‌گیری هموارتر شده است و زیر خط فرضی ظرفیت درمانی قرار گرفته است.

## ۳- مدل هموارکردن منحنی همه‌گیری

### ۳-۱- معیارهای مرکزیت

معیارهای مرکزیت (centrality measures) کمیت‌هایی برای مشخص کردن میزان اهمیت نودها در شبکه هستند. به عبارت دیگر این معیارها مهم بودن یا مهم نبودن یک نود را برای یک ماربرد و هدف خاص در شبکه‌های انسانی-اجتماعی مشخص می کنند. معیارهای مختلفی برای نشان دادن میزان اهمیت نودها براساس موقعیت ساختاری نود در گراف شبکه ارایه شده است که در ادامه لیستی از این معیارها آورده شده است. در این مقاله این معیارها برای انتخاب نودهای هدف برای ایزوله کردن (یا قرنطینه کردن) مورد استفاده قرار میگیرند. تعدادی از این معیارها در مرجع [۳] و مقاله مروری [۲۳] لیست شده‌اند.

**درجه (degree):** برابر است با تعداد یال‌هایی از گراف که به یک نود متصل اند یا به عبارتی تعداد همسایه‌ها

بین آن نود و تمام نودهای قابل دسترس به آن بدست می‌آید. این مسیرها وزندار می‌باشند به طوری که وزن مسیرهایی که همسایه مجاور را به آن نود وصل می‌کنند بزرگتر از وزن مسیرهایی است که همسایه‌های دورتر را به آن نود وصل می‌کنند [۲۵].

رتبه پیج (Page rank):
در این معیار ابتدا به همه نندها ارزش یکسانی داده می شود و سپس هر نود ارزش خود را به نسبت مساوی بین همسایگانش تقسیم میکند و ارزش جدید هر نود برابر است با مجموع امتیازاتی که نودهای همسایه به آن میدهند. این پروسه تا زمانی که تمام امتیازات نودها همگرا شوند ادامه پیدا می کند و در نهایت امتیاز هر نود برابر است با رتبه پیج آن نود. این معیار در واقع هم درجه نود و هم یکتا بودن لینکهای متصل به آن نود را ترتیب اثر می دهد. به عبارت دیگر نود i دارای رتبه پیج بالایی است اگر نودهای همسایه اش j دارای مرکزیت بالایی باشند و نود i یکی از معدود همسایه های نود j باشد. [۲۶]

نیروی مورد انتظار (expected force): این معیار که براساس آنتروپی تعریف شده به صورت خاص برای مساله اپیدمی ارایه شده است [۲۷]. ارزش یک نود طبق این معیار براساس تعداد موردانتظار نیروی انتقال بیماری (force of infection) بعد از دو پله اثرگذاری و انتقال بیماری از نود موردنظر (آلوده) تعریف می شود. در این معیار درجه یک نود نیز دخیل می شود و بطور کلی یک نود با درجه کم ممکن است ارزش بالایی داشته باشد در صورتی که با نودهای با درجه بالا در ارتباط باشد.

این معیارها را میتوان به صورت کلی یا به صورت محلی در هر نود محاسبه کرد (بطور مثال با استفاده از استراتژی رندوم [۲۸]). پس از تعیین مرکزیت نودها هدف بعدی ایزوله کردن نودهای با مرکزیت بالا براساس این معیارها می باشد.

## ۳-۲- ایزوله کردن نودها

برای هموار کردن منحنی همه گیری باید نودهای با اهمیت بالا از لحاظ اپیدمیولوژیکی را ایزوله یا قرنطینه کرد تا نرخ انتقال بیماری را در شبکه با افزایش طول مسیر انتقال بیماری کند کرد. به عبارت دیگر با قرنطینه کردن نودهایی که دارای مرکزیت بالایی هستند طول کوتاه ترین مسیر بین نودها در شبکه افزایش می یابد و مسیر انتقال بیماری بین نودها طولانی تر شده و در نتیجه نمودار همه گیری هموارتر می شود. مثالی از هموارتر کردن نمودار همه گیری با ایزوله کردن نودها پیشتر در شکل (۳) ارایه شد. در این بخش برای ایزوله کردن نودها انواع معیارهای مرکزیت مورد بررسی قرار می گیرند. به این ترتیب که تعداد معینی از نودهای با مرکزیت بالا ایزوله شده و لینکهای ارتباطی آنها با نودهای دیگر قطع می شود. سپس اثر ایزوله کردن روی منحنی همه گیری با شبیه سازی مونته کارلو (Monte-Carlo) بررسی می شود. برای این منظور شبکه های بی مقیاس با تعداد ۲۰۰ نود براساس مدل باراباشی-آلبرت و مدل هولم-کیم با ضرایب خوشه بندی ( clustering coefficient) مختلف درنظر گرفته میشود. نمودار همه گیری برای ۱۰ شبکه بیمقیاس میانگین گیری مونته کارلو می شود. این نمودار در شکل (۴) آمده است. توجه شود که در این مرحله هنوز ایزوله کردن نودها صورت نگرفته است.

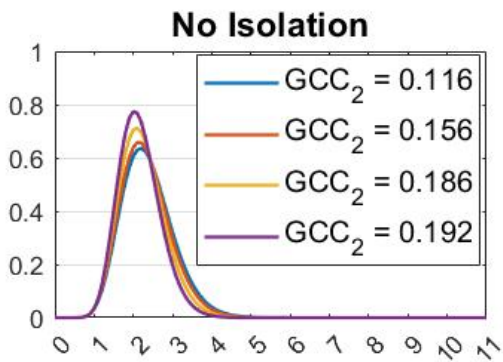

شکل ۴- شبیه سازی مونته کارلو نمودار همه گیری برای ضرایب خوشه بندی سراسری (GCC) مختلف

در مرحله بعد تعداد ۳٪ نودها براساس معیارهای مرکزیت مختلف ایزوله می شوند و نمودار همه گیری برای همان ۱۰ شبکه بیمقیاس میانگین گیری مونته کارلو می شود. به این ترتیب نمودار همه گیری هموارشده (بعد از ایزوله کردن) براساس معیار درجه در شکل (۵)، براساس معیار

نزدیکی در شکل (۶)، براساس معیار بینابینی در شکل (۷)، براساس معیار بوناچیچ در شکل (۸)، براساس معیار کاتز در شکل (۹)، براساس معیار رتبه پیج در شکل (۱۰)، و براساس معیار نیروی مورد انتظار در شکل (۱۱) آمده است. توجه کنید که نمودارها برای شبکه های بیمقیاس با خوشه بندی های مختلف تکرار شده است تا اثر خوشه بندی هم در ایزوله کردن مشخص شود.

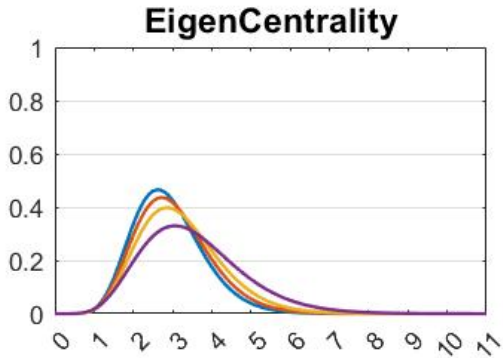

شکل ۸- هموارکردن نمودار همه گیری با ایزوله کردن ۳٪ نودهای مهم براساس معیار بوناچیچ

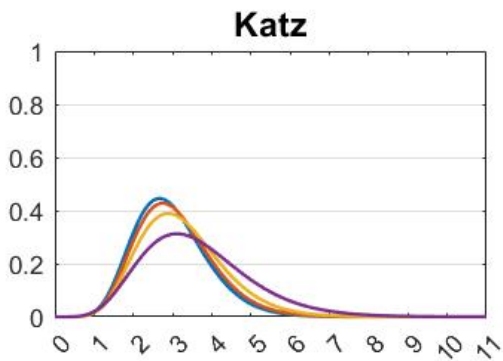

شکل ۹- هموارکردن نمودار همه گیری با ایزوله کردن ۳٪ نودهای مهم براساس معیار کاتز

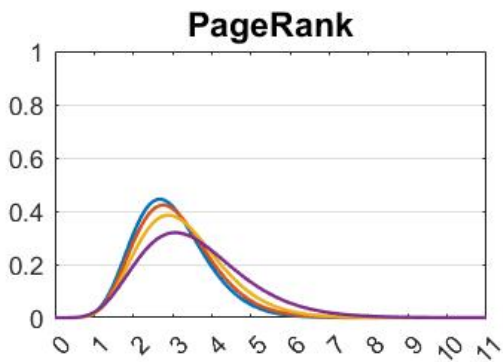

شکل ۱۰- هموارکردن نمودار همه گیری با ایزوله کردن ۳٪ نودهای مهم براساس معیار رتبه پیج

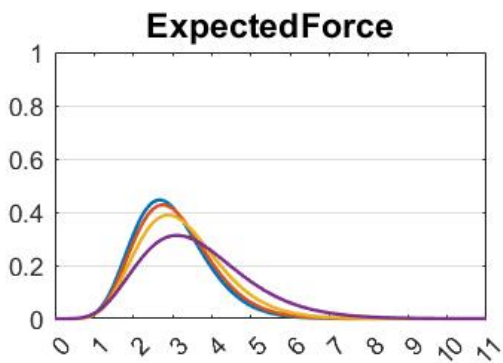

شکل ۱۱- هموارکردن نمودار همه گیری با ایزوله کردن ۳٪ نودهای مهم براساس معیار نیروی موردانتظار

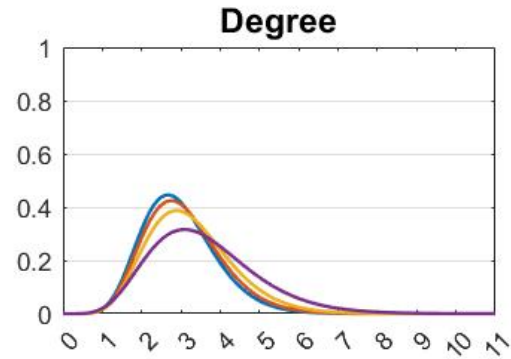

شکل ۵- هموارکردن نمودار همه گیری با ایزوله کردن ۳٪ نودهای مهم براساس معیار درجه

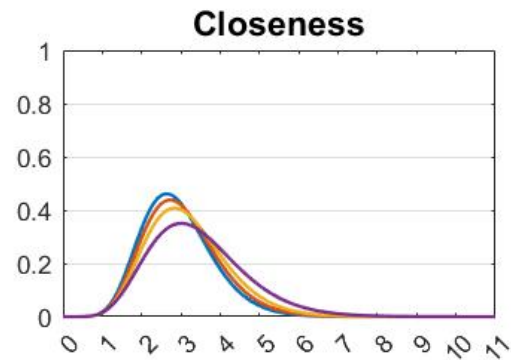

شکل ۶- هموارکردن نمودار همه گیری با ایزوله کردن ۳٪ نودهای مهم براساس معیار نزدیکی

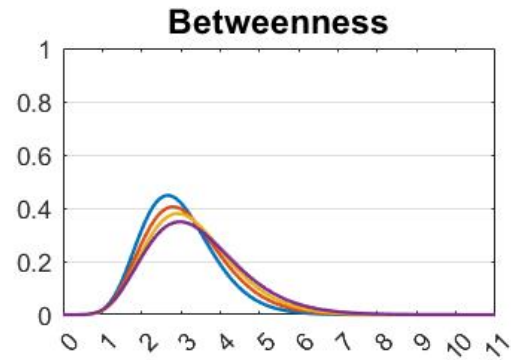

شکل ۷- هموارکردن نمودار همه گیری با ایزوله کردن ۳٪ نودهای مهم براساس معیار بینابینی

همانطور که مشاهده میشود در مقایسه شکلهای (۵) تا (۱۱) که نمودار را بعد از ایزوله کردن نشان میدهند با شکل (۴) که در آن هیچ ایزوله کردنی صورت نگرفته نمودارهای همه گیری تا حد محسوسی هموار شده اند. نکته جالب اینکه نمودارهای مربوط به شبکه های بیمقیاس با خوشه بندی بالاتر به نسبت بیشتری هموار شده اند. همچنین هر دو دسته معیارهای بر مبنای درجه و بر مبنای مسیر معیارهای مناسبی برای انتخاب نودهای هدف برای ایزوله کردن (قرنطینه) هستند.

برای ارایه یک شاخص برای مقایسه منحنیها مقدار پیک منحنی پیش از نرمالیزه کردن در جدول ۱ آمده است. در این جدول اعداد نشان داده شده مقدار بیشترین تعداد بیمار قبل و بعد از ایزوله کردن (یا واکسیناسیون) نودهای هدف بر اساس معیارهای مختلف مرکزیت را نشان میدهد. مقدار نرمالیزه شده این مقادیر (پس از تقسیم بر تعداد کل مسیرهای انتقال بیماری) به شکل منحنی های توزیع احتمالاتی در شکلهای ۴ تا ۱۱ به نمایش گذاشته شده است. همانطور که از جدول ۱ مشخص است مقدار پیک بیماری تا حد محسوسی بعد از ۳٪ ایزولاسیون نودها کاهش پیدا کرده است و این کاهش برای شبکه های با خوشه بندی بالاتر محسوستر و مشخص تر است.

جدول ۱- پیک بیماری بدون ایزوله کردن و با روشهای ایزوله کردن متفاوت برای مقادیر مختلف خوشه بندی

| GCC | 0.116 | 0.156 | 0.186 | 0.192 |
|---|---|---|---|---|
| None | ۶۱۳۷ | ۶۷۱۶ | ۷۷۳۱ | ۸۵۸۲ |
| Deg | ۴۱۲۴ | ۳۴۱۳ | ۲۹۳۲ | ۲۴۵۶ |
| Clos | ۴۱۱۳ | ۳۶۳۱ | ۳۰۸۸ | ۲۶۶۲ |
| Bet | ۴۱۳۹ | ۳۶۰۱ | ۳۰۹۰ | ۲۳۴۲ |
| Eig | ۴۱۴۷ | ۳۶۲۱ | ۳۰۱۵ | ۲۵۹۸ |
| Katz | ۴۱۲۵ | ۳۴۲۶ | ۲۹۴۳ | ۲۴۹۶ |
| Page | ۴۱۱۷ | ۳۴۰۰ | ۲۹۲۵ | ۲۴۱۵ |
| Exf | ۴۱۲۵ | ۳۴۲۳ | ۲۹۴۷ | ۲۴۹۹ |

توجه شود که ایزوله کردن نود در گراف به معنی حذف راههای انتقال بیماری از این نود به نودهای همسایه است. در دنیای واقعی در این مورد میتوان از واکسیناسیون هدفمند نیز استفاده کرد، به این معنی که با واکسیناسیون فرد لینکهای انتقال بیماری به همسایه ها و اطرافیان از فرد واکسینه شده (به جای ایزوله کردن و قرنطینه) تا حد خوبی حذف می شود. به عبارت دیگر، بطور مثال درمورد نود ایزوله شده در گراف با رعایت کامل پروتکلهای بهداشتی، فاصله گذاری اجتماعی و یا واکسیناسیون راههای انتقال بیماری به اطرافیان (نودهای همسایه در گراف) حذف شده اند و امکان انتقال بیماری از آن نود به نودهای دیگر همسایه اش وجود ندارد. توجه شود که این مقاله صرفا به مساله بیماری کرونا نمیپردازد بلکه هر نوع اپیدمی را شامل می شود.

### ۳-۳- نوآوری ها

بطور کلی نوآوریهای این مدل در ارایه یک مدل ریاضی با استفاده از تیوری گرافها برای گسترش یک بیماری در شبکه های انسانی-اجتماعی و مساله هموار کردن منحنی همه گیری در شبکه است. این مدل امکان بررسی گسترش انواع بیماری ها در انواع شبکه ها را میدهد و یک مدل بسیار کلی است. نوآوری بعدی در ارایه یک روش برای هموار کردن منحنی همه گیری مدل شده است. به این ترتیب که با ایزوله کردن تعدادی از نودها و حذف لینکهای این نودها مساله قرنطینه اشخاص یا واکسیناسیون اشخاص شبیه سازی می شود. با استفاده از این شبیه سازی ها می توان انواع روشها را برای انتخاب نودهای هدف ایزولاسیون را بررسی کرد. روش پیشنهادی در این مقاله استفاده از معیارهای مرکزیت است که نودها را براساس اهمیت رتبه بندی می کند و سپس نودهای با اهمیت بالا ایزوله می شوند. این نوآوری امکان بررسی انواع معیارهای مرکزیت را فراهم هم می کند. نو آوری بعدی امکان بررسی شبکه ها با خوشه بندی های متفاوت است. به این ترتیب که با تغییر خوشه بندی شبکه مساله هموارکردن منحنی اپیدمی با استفاده از معیارهای مرکزیت مختلف بررسی می شود. نتیجه جالب این که شبکه های با خوشه بندی بالاتر با استفاده از این روش ایزوله کردن نودهای با مرکزیت بالا نتیجه بهتر و منحنی اپیدمی هموارتری می دهد.

### ۴- نتیجه گیری

در این تحقیق هموارکردن نمودار همه گیری بیماری با روش ایزوله کردن درصد کمی از نودها در شبکه های بیمقیاس مورد بررسی قرار گرفته است. ایزوله کردن یک نود در دنیای واقعی نمایانگر قرنطینه کردن یک شخص یا حتی واکسیناسیون یک فرد برای قطع انتقال بیماری از شخص موردنظر به اشخاص دیگر (نودهای همسایه) است. نتایج این تحقیق نشان می دهد با قرنطینه (ایزوله کردن) درصد کمی از افراد خاص (نودهای با مرکزیت بالا در شبکه) امکان هموارکردن نمودار همه گیری به میزان بالایی در جامعه وجود دارد. همچنین نمودارها نشان میدهند که ایزوله کردن نودها در شبکه های با خوشه بندی بالاتر اثرات موثرتری دارد و باعث هموارشدن بیشتر منحنی همه گیری می شود.

اهداف و تاثیر نتایج این پژوهش به طور خلاصه راهکارهای بسیاری را برای قرنطینه هدفمند افراد در شبکه های انسانی-اجتماعی فراهم می کند. بطور مثال امکان سنجی واکسیناسیون و ایزوله کردن هدفمندتر با درنظر گرفتن موقعیت (مرکزیت) ارتباطی-اجتماعی اشخاص که بالتبع موجب کاهش نرخ انتقال در موجها و پیکهای اپیدمی خواهد شد. به عبارت دیگر، هدف نهایی ارایه راهکارهایی درمانی یا غیر درمانی بر اساس شبکه که باعث کم کردن بار درمانی در پیک بیماری و کاهش بار درمانی بر بیمارستانها و مراکز خدمات بهداشتی با قرنطینه و واکسیناسیون "هدفمند" میباشد. منظور از قرنطینه هدفمند انتخاب گروه هدف برای ایزوله کردن (نودهای با مرکزیت بالا) می باشد. امکان سنجی جایگزین کردن قرنطینه عمومی غیر هدفمند با "قرنطینه و ایزوله کردن هدفمند فردی" به منظور کاهش بار اقتصادی بر جامعه و کسب و کارها بر اساس تعاملات مدل شده با شبکه انسانی-اجتماعی از دیگر نتایج و تاثیرات عمومی این تحقیق است.

به عنوان مسیر تحقیقات آینده میتوان از انواع مختلف گرافهای موجود برای مدل کردن شبکه اجتماعی و انسانی استفاده کرد یا در صورت امکان گرافهای واقعی را در نظر گرفت. بطور مثال میتوان به جای مدل بی مقیاس از مدل جهان کوچک (small-world) استفاده کرد. وزندار کردن نودها (یا اشخاص) باتوجه به اهمیت شغلی آنها روش دیگری است که میتواند مورد استفاده قرار گیرد. به این معنی که ضمن در نظر گرفتن وزن هایی متفاوت برای نودها ، عدم امکان حذف برخی از نودها که مشاغل مهمی هستند و یا حذف آنها هزینه بر است را در نظر گرفت. یا به طور مثال در صورتی که با تعداد محدودی واکسن مواجه بودیم با وزندار در نظر گرفتن نودها این تعداد واکسن را برای واکسیناسیون افراد مهمتر در جامعه هدف استفاده کرد. به این ترتیب مساله به یک مساله بهینه سازی تبدیل می شود که باید به شکل بهینه ای نودها با توجه به وزن آنها حذف (ایزوله) و یا واکسینه شوند.

# مراجع

# Epidemic modeling and flattening the infection curve in social networks

**Mohammadreza Doostmohammadian[1], Soraya Doustmohamadian[2], Najmeh Doostmohammadian[3], Azam Doustmohammadian[4], Houman Zarrabi[5], Hamid R. Rabiee[6]**

1. Faculty of Mechanical Engineering, Semnan University
2. Semnan University of Medical Sciences
3. Semnan University of Medical Sciences
4. Gastrointestinal and Liver Diseases Research Center, Iran University of Medical Sciences
5. Iran Telecom Research Center (ITRC)
6. Faculty of Computer Engineering, Sharif University of Technology

*Corresponding Author: Mohammadreza Doostmohammadian, doost@semnan.ac.ir

**ABSTRACT**

The main goal of this paper is to model the epidemic and flattening the infection curve of the social networks. Flattening the infection curve implies slowing down the spread of the disease and reducing the infection rate via social-distancing, isolation (quarantine) and vaccination. The nan-pharmaceutical methods are a much simpler and efficient way to control the spread of epidemic and infection rate. By specifying a target group with high centrality for isolation and quarantine one can reach a much flatter infection curve (related to Corona for example) without adding extra costs to health services. The aim of this research is, first, modeling the epidemic and, then, giving strategies and structural algorithms for targeted vaccination or targeted non-pharmaceutical methods for reducing the peak of the viral disease and flattening the infection curve. These methods are more efficient for nan-pharmaceutical interventions as finding the target quarantine group flattens the infection curve much easier. For this purpose, a few number of particular nodes with high centrality are isolated and the infection curve is analyzed. Our research shows meaningful results for flattening the infection curve only by isolating a few number of targeted nodes in the social network. The proposed methods are independent of the type of the disease and are effective for any viral disease, e.g., Covid-19.